\documentclass[journal]{IEEEtran}

\ifCLASSINFOpdf

\else

\fi

\usepackage{cite}
\usepackage{graphicx}
\usepackage{amsmath}
\usepackage{subcaption}
\usepackage{flushend}
\usepackage{algpseudocode}
\usepackage[linesnumbered,ruled,vlined,commentsnumbered]{algorithm2e}
\usepackage{lipsum}
\usepackage[font=footnotesize]{caption}
\usepackage{graphicx}
\usepackage{subcaption}
\usepackage{amssymb}
\usepackage{gensymb}
\usepackage{wasysym}
\usepackage{textcomp}
\usepackage{color}

\usepackage{mathtools}
\DeclarePairedDelimiterX{\norm}[1]{\lVert}{\rVert}{#1}

\makeatletter
\renewcommand{\@algocf@capt@plain}{above}% formerly {bottom}

\begin{document}

\title{Distributed Dynamic Pricing in Peer-to-Peer Transactive Energy Systems in Smart Grid}
\author{Md Habib~Ullah, Anas~Alseyat, and Jae-Do~Park\\
\IEEEauthorblockA{Dept. of Electrical Engineering, University of Colorado Denver\\
Email: md.ullah@ucdenver.edu, anas.alseyat@ucdenver.edu, jaedo.park@ucdenver.edu}
}

%\author{Md Habib~Ullah,~\IEEEmembership{Student Member,~IEEE,}
%Anas Alseyat,
%Jae-Do~Park,~\IEEEmembership{Senior Member,~IEEE}\\
%\IEEEauthorblockA{Dept. of Electrical Engineering, University of Colorado Denver\\
%Email: md.ullah@ucdenver.edu, jaedo.park@ucdenver.edu}
%}
\maketitle
\footnote{\textcolor{blue}{The final published paper is copyrighted by IEEE and should be cited as: M. H. Ullah, A. Alseyast, and J.-D. Park, ``Distributed dynamic pricing in peer-to-peer transactive energy systems in smart grid,” in \textit{IEEE Power \& Energy Society General Meeting (PESGM)}, 2020, pp. 1–5. ``Personal use of this material may be permitted and permission from IEEE must be obtained for all other uses."}}\begin{abstract}
The rapid growth of proactive consumers with distributed power generation and storage capacity, empowered by Internet of Things (IoT) devices, is transforming modern power markets into an independent, flexible, and distributed structure. In particular, the recent trend is peer-to-peer (P2P) transactive energy systems, wherein the traditional consumers became prosumers (producer+consumer) and can maximize their energy utilization by sharing with neighbors without any conventional intermediary intervention in the transactions. However, the competitive dynamic energy pricing scheme is inevitable in such systems to make the optimal decision. It is very challenging when the prosumers have limited access to the fellow prosumer's system information (i.e., load profile, generation, and so on). This paper presents a privacy-preserving distributed dynamic pricing strategy for P2P transactive energy systems in the smart grid using Fast Alternating Direction Method of Multipliers (F-ADMM) algorithm. The result shows that the algorithm converges very fast and facilitates easy implementation. Moreover, a closed-form solution for a P2P transactive energy system was presented, which accelerate the overall computation time.
\end{abstract}
%
%\vspace{-2pt}

\markboth{This paper has been accepted and presented in \underline{\textit{IEEE PESGM 2020}} conference and is in the process for publication.}%
{}

\IEEEpeerreviewmaketitle

\section{Introduction}
\IEEEPARstart{I}{t} is expected that the global market for rooftop photovoltaic (PV) panels will worth of about \$33 billion by 2022, and the increase in the adoption of residential energy storage systems complements the shift towards PV even further whose capacity is predicted to grow more than 3.7 GW by 2025 \cite{peck2017energy}. These additional energy resources at the edge of the grid are expected to be utilized more efficiently not only to manage the energy demand but also to enable a significant penetration of renewable energy into the grid. In such scenarios, it is of utmost importance for the edge users with generating assets to be incorporated into the energy market as prosumers (producer+consumer).

The role of prosumers is well recognized in the deregulated energy market. The two dominant schemes for compensating the prosumers for the energy they feed back into the grid are net metering and feed-in tariff (FIT) programs \cite{peck2017energy}. In net metering, tariffs are designed for the prosumers to receive an incentive on their utility bills using electricity at certain times by offsetting their use of electricity from the grid at other times \cite{darghouth2011impact}. FITs are the most widely used policy throughout the world for accelerating renewable energy (RE) deployment. With FITs, the prosumers with roof-top solar panels can sell their excess solar energy to the grid and can buy energy from the grid in case of any energy deficiency \cite{couture2010policymaker}. However, the high penetration of PVs in the grids raises stability and reliability issues; hence, the local governments in many countries have limited the PV export to the grid \cite{tushar2018peer}. Moreover, the prosumers are getting paid at a fixed rate in both the schemes \cite{peck2017energy}. Typically, the prosumers buy energy from the grid at a high price and get paid low prices when they sell; therefore, they lose potential benefits.

In this context, peer-to-peer (P2P) energy sharing concept has emerged in the area of the distribution networks \cite{morstyn2018using}. Unlike the traditional systems, the P2P scheme enables the prosumers to participate in a local energy arbitrage with the neighboring prosumers. Currently, the P2P distributed market platforms is possible due to advances in information and communication technology, and distributed ledger technologies (DLTs) such as blockchain, which support transparent and decentralized transactions \cite{guerrero2018decentralized}. However, efficient energy pricing in such systems is an important task.

In recent literature, various approaches have been presented for P2P energy trading in distribution systems. For instance, an auction-based P2P energy trading model is introduced for prosumer-centric community microgrids using unified and identified pricing strategies \cite{wu2018user}. In \cite{long2017peer}, a P2P energy trading paradigm has been presented with various pricing strategies such as bill sharing, mid-market rate, and auction-based, and various scenarios were compared. In addition to the mid-market rate, a canonical coalition game-theoretic model has been adopted for prosumer-centric microgrids in \cite{tushar2018peer}. In addition, various trading approaches such as power-based tariff \cite{narayanan2018economic}, auction-based market-clearing framework \cite{khorasany2017auction,peerkhorasany2017peer} are presented for P2P transactive energy systems. Further, a two-stage bidding strategy for P2P energy trading in residential microgrids is proposed \cite{zhang2019two}. In \cite{liu2015energy}, a P2P energy sharing model is presented for a distribution system. A single-layered P2P energy trading system is proposed for heterogeneous small-scale DERs \cite{yoo2017peer}. However, these approaches are in a centralized manner and require a central coordinator to perform the trading, which may violate the privacy. Moreover, a centralized system is prone to a single point of failure. In this context, Relaxed Consensus+Innovation-based distributed pricing strategies have been proposed \cite{moret2018negotiation,sorin2018consensus}. However, the consensus-based approaches suffer from slow convergence rate. In improvement, an alternating direction method of multipliers (ADMM) method for prosumer's preference-driven P2P energy trading model is presented \cite{morstyn2018multi}. Additionally, game-theoretic approaches have been presented in \cite{le2019peer,anoh2019energy}.

In this paper, a F-ADMM-based distributed dynamic energy pricing scheme is proposed for P2P trasactive energy systems. The proposed algorithm is scalable, requires only buyer-to-seller P2P communications, and reduces information sharing to settle down the market. A closed form solution is also derived to reduce computation effort. Moreover, the proposed approach preserves privacy for the trading participants.

The rest of this paper is organized as follows: Section \ref{Preliminaries} presents the description of a prosumer-based distribution system, mathematical modeling of the prosumers utility function, and the centralized optimization problem formulation. The distributed closed-form solution is presented in Section \ref{Distributed Solution}. Section \ref{Result} shows the simulation result and analysis followed by Section \ref{conclusion} that concludes the paper.

\section{Preliminaries and Problem Formulation}\label{Preliminaries}
\begin{figure}[t]
    \centering
    \includegraphics[scale=0.6]{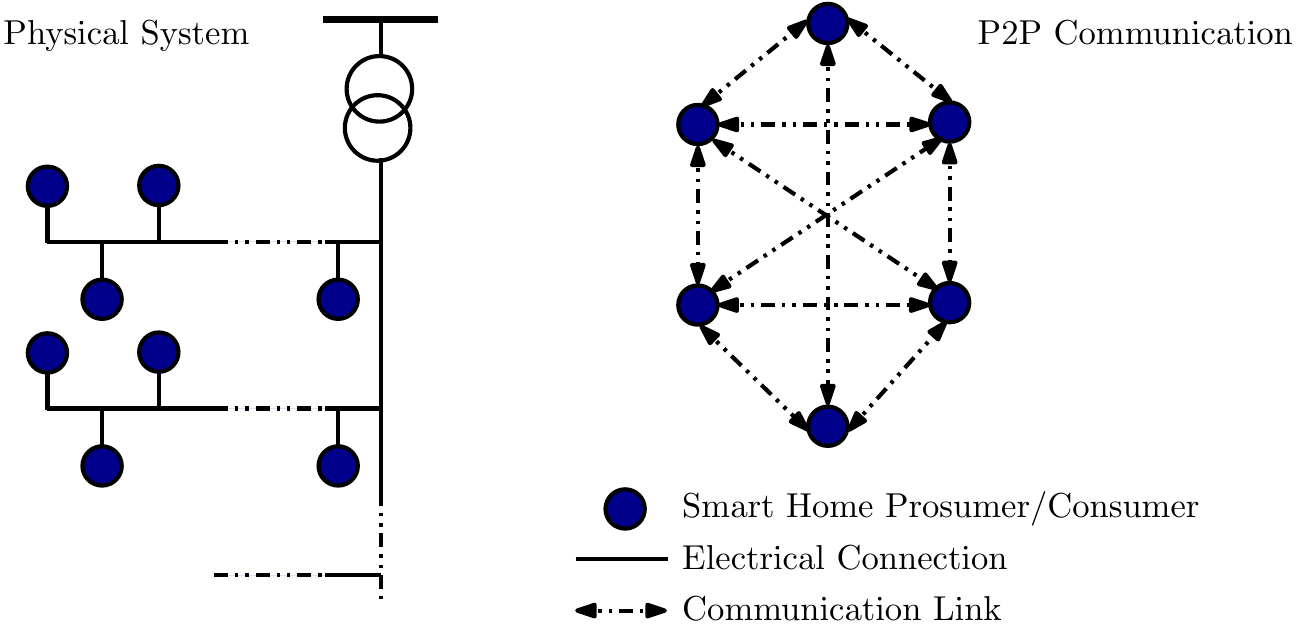}
    \caption{Conceptual distribution system with peer-to-peer (P2P) connection.}
    \label{fig. 1}
    \vspace{-6pt}
    \end{figure}

Consider a smart distribution network shown in the Fig. \ref{fig. 1} comprises of a set of prosumers be $\mathcal{N} = \{1, 2, 3, ... N\}$, where $N = |\mathcal{N}|$ gives the total number of prosumers in the system. Each prosumer has roof top PV panels installed and connected to the network through a smart meter. In each time step, the prosumers declare themselves as a producer or a consumer based on their net load. Multi-agent systems incorporated with blockhain and other distributed ledger technologies can be utilized to ensure system transparency. Let there $\{\mathcal{P} = \{\mathcal{P}_x : \forall x \in \mathcal{N} \} \}$ producers and $\{\mathcal{C} = \{\mathcal{C}_y : \forall y \in \mathcal{N}_s \} \}$ consumers; $\mathcal{N}_s \in \mathcal{N}$ is the set of indices of consumers that buy energy from producer $y$. However, the subscript $x$ or $y$ can be replaced with appropriate producer or consumer index. It should be noted that $\mathcal{N} = \mathcal{P} \cup \mathcal{C} $ and $\mathcal{P} \cap \mathcal{C} = \emptyset$. Both the $\mathcal{P}$ and $\mathcal{C}$ are connected in the physical power distribution network. The P2P connectivity can be denoted using an energy trading graph $\mathcal{G} = (\mathcal{N}, \mathcal{E})$ with a set of edges $\mathcal{E} \subseteq N \times N$.
%Each edge represents a communication link between two peers $n$ and $m$ with weight $\mathcal{A}_{n,m}$.
In this paper, the connections between producer-to-consumer are required, not between producer-to-producer or consumer-to-consumer.
%, which implies $\mathcal{A}_{n,m} = 1$, $\mathcal{A}_{n,n} = 0$, and $\mathcal{A}_{m,m} = 0$.
However, the energy storage system was not considered in this paper. Also, the energy is called as power, considering an hour-based operation.
%
%\subsection{P2P Trading Problem Formulation}
\subsection{Prosumer Utility Function}
The utility function of a prosumer known as cost function is typically convex, which quantifies their satisfaction level on using a certain amount of power. There are logarithmic and quadratic utility functions frequently used for modeling the prosumer utility function. Without losing the generality, a quadratic utility function is considered in this paper as \cite{sorin2018consensus}:
\begin{equation}\label{Eq.1}
c_n(\hat{p}_n) = \alpha_n \hat{p}_n^2 + \beta_n\hat{p}_n + \gamma_n,
\end{equation}
where $\alpha_n>0$ and $\beta_n, \gamma_n \geq 0$ are the utility function coefficients. The power $\hat{p}_n$ is positive and negative for the producers and consumers, respectively.
%It should be noted that subscript $x \in \{n, m\}$, where $n \in \mathcal{P}$ and $m \in \mathcal{C}$.
The power set point for each prosumer is constrained by its minimum, $\hat{p}_n^{-}$ and maximum, $\hat{p}_n^{+}$  limits as:
\begin{equation}\label{Eq.2}
\hat{p}_n^{-} \leq \hat{p}_n \leq  \hat{p}_n^{+}.
\end{equation}
\subsection{Bilateral Trading Problem Formulation}
Let $\boldsymbol p_{n} = \{ p_{n,m},  \forall n \in \mathcal{N}, \forall m \in \mathcal{N} \backslash n \}$
%or, $\boldsymbol p_{n} = \{ p_{n,m},  \forall m \in \mathcal{P}, \forall n \in \mathcal{C} \}$
is the traded power between the prosumer $n$ and its neighbors $m \in \mathcal{N} \backslash n$. To avoid energy sharing between producer-to-producer or consumer to consumer, it should be considered that if $n$ is producer then $m$ is consumer, and vice-versa. The traded power, $p_{n,m}$ is positive and negative when prosumer $n$ sells and buys power to/from $m$, respectively. Therefore, $p_{n,m}$ and $p_{m,n}$ are equal valued but of opposite polarity and each prosumer should satisfy the following reciprocity coupled constraint:
\begin{equation}\label{Eq.3}
p_{n,m} + p_{m,n} = 0.
\end{equation}
%
%\begin{equation}\label{Eq.1}
% P_{m} = \sum_{n\in \mathcal{P}} p_{m,n}
%\end{equation}
%
To model the P2P trading scheme, the power set point $\hat{p}_n$ of each prosumer can be given as the sum of P2P traded quantities as:
\begin{equation}\label{Eq.4}
 \hat{p}_n = \sum_{m \in \mathcal{N} \backslash n} p_{n,m}.
\end{equation}
Let the producer $n \in \mathcal{P}$ and consumer $m \in \mathcal{C}$ privately negotiate the price of the traded power $p_{n,m}$ as $\pi_{n,m}$. The following constraint should be satisfied considering the balance of payments in each pair of producer and consumer as:
\begin{equation}\label{Eq.5}
\pi_{n,m} = \pi_{m,n}.
\end{equation}
It should be noted that, $p_{n,m}$ and $\pi_{n,m}$ are the traded power and unit price determined by $n$; and similarly, $p_{m,n}$ and $\pi_{m,n}$ are the traded power and price determined by $m$.

%\subsection{Social Welfare}
%Prosumer's social welfare is commonly known as the benefit derived from the market on which they trade power with the grid, which can be determined as \cite{hug2015consensus+}:
%
%\begin{equation}\label{Eq.6}
%\mathcal{S}_n = \boldsymbol p_n^\top \boldsymbol \pi_n - c_n(\hat{p}_n),
%\end{equation}
%
%where $\boldsymbol \pi_n = \{ \pi_{n,m}, m \in \mathcal{N} \backslash n \}$ is the price vector determined by prosumer $n$, and $\boldsymbol p_n^\top$ is the transpose of the vector of traded power between $n$ and $m$.

\subsection{Centralized Optimization} \label{centralized optimization}
The centralized trading problem can be formulated to minimize the combined utility function %or maximize the social welfare of
the prosumers. Therefore, the overall trading problem can be written as:
\begin{subequations}\label{Eq.7}
\begin{align}
\min_{{\hat{p}_n}} \quad
& \sum_{n\in \mathcal{N}}{c_n(\hat{p}_n)}  \label{Eq.7_1}\\
\text{s.t.}\quad
& \hat{p}_n^{-} \leq \hat{p}_n \leq  \hat{p}_n^{+}, \;\;\;\;  \forall n\in \mathcal{N}, \label{Eq.7_2}  \\[2mm]
& p_{n,m} + p_{m,n} = 0, \;\;\;\; \forall n \in \mathcal{N}, \;\;\;\; \forall m \in \mathcal{N} \backslash n,  \label{Eq.7_3}  \\[2mm]
& p_{n,m} \geq 0, \;\;\;\;\;\;\;\;\;\;\;\;\;\;\;\; \forall n \in \mathcal{P},\;\;\;\; \forall m \in \mathcal{N} \backslash n, \label{Eq.7_4}\\[2mm]
& p_{n,m} \leq 0, \;\;\;\;\;\;\;\;\;\;\;\;\;\;\;\; \forall n \in \mathcal{C},\;\;\;\; \forall m \in \mathcal{N} \backslash n. \label{Eq.7_5}
\end{align}
\end{subequations}
The optimization problem (\ref{Eq.7}) can readily be solved using a conventional solver. However, a central coordinator is required, and prosumers may need to share their utility function coefficients and energy generation and consumption profile, which perhaps raise privacy issues in the systems because the prosumers may not be willing to disclose this information. Considering that, a privacy-preserving distributed optimization problem is formulated in the following section, and the closed-form solution is derived afterwards.

\section{Distributed Solution in P2P Energy Trading}\label{Distributed Solution}
In this section, the centralized trading problem is formulated into sub-problem and solved using a distributed iterative approach with closed-form solution.

\subsection{ADMM-based Solution}
In (\ref{Eq.7_3}), it can be seen that the power $p_{n,m}$ is coupled between prosumer $n$ and $m$, which is in multi-block structure. This can be modified into a two-block subproblem and can be solved using standard ADMM introducing auxiliary variables $\sigma$.

In this paper, F-ADMM \cite{goldstein2014fast} algorithm was adopted instead of standard ADMM to accelerate the convergence speed. However, the coupled constraint (\ref{Eq.7_3}) can be reformulated as:
\begin{equation}\label{Eq.8}
\sigma_{n,m} = p_{n,m}, \;\;\; \forall n \in \mathcal{N}, \forall m \in \mathcal{N} \backslash n,
\end{equation}
\begin{equation}\label{Eq.9}
\sigma_{n,m} + \sigma_{m,n} = 0, \;\;\; \forall n \in \mathcal{N}, \forall m \in \mathcal{N} \backslash n,
\end{equation}
\begin{equation}\label{Eq.10}
\sigma_{m,n} = p_{m,n}, \;\;\; \forall n \in \mathcal{N}, \forall m \in \mathcal{N} \backslash n.
\end{equation}
Defining $\boldsymbol \pi_n = \{ \pi_{n,m}\}$ as the dual variables associated with (\ref{Eq.8}), the augmented Lagrangian for (\ref{Eq.7}) can be written as:
\begin{equation}\label{Eq.11}
\begin{split}
& \mathcal{L}(\hat{\boldsymbol p_n}, \boldsymbol \pi_{n,m}, \boldsymbol \sigma_{n,m}) = \sum_{n\in \mathcal{N}} \bigg[ {c_n(\hat{p}_n)} \\
& + \sum_{m \in \mathcal{N} \backslash n} \bigg\{ \pi_{n,m} \big(\sigma_{n,m} - p_{n,m} \big)  + \frac{\kappa}{2} \big(\sigma_{n,m} - p_{n,m} \big)^2\bigg \} \bigg],
\end{split}
\end{equation}
where, $\kappa$ is a well-defined penalty parameter for the constraint (\ref{Eq.8}).
The ADMM is an iterative approach; it involves solving three different problems in each iteration. Specifically, the first problem ($\boldsymbol P_1$) associates with solving the local problem based on the current dual variable $\boldsymbol \pi_n$, and auxiliary variable $\boldsymbol \sigma_n = \{ \sigma_{n,m}, m \in \mathcal{N} \backslash n \}$. At $i^{th}$ iteration, each prosumer solves the following local optimization problem to determine the traded power $p_{n,m}$ with given $\pi_{n,m}^i$ and $\sigma_{n,m}^i$ as:
%
%$\boldsymbol P_1$: The local optimization problem can be written as:
%
\begin{subequations}\label{Eq.12}
\begin{align}
\begin{split}
\min_{{p_{n,m}}} \quad
&  {c_n(\hat{p}_n)} + \sum_{m \in \mathcal{N} \backslash n} \bigg\{ \pi_{n,m}^i \big(\sigma_{n,m}^i - p_{n,m} \big)  \\
& + \frac{\kappa}{2}  \big(\sigma_{n,m}^i - p_{n,m} \big)^2 \bigg \}, \label{Eq.12_1}\\
\end{split}\\
\text{s.t.}\quad
& \hat{p}_n^{-} \leq \hat{p}_n \leq  \hat{p}_n^{+}, \;\;\;\;  \forall n\in \mathcal{N}, \label{Eq.12_2}  \\[2mm]
& p_{n,m} \geq 0, \;\;\;\;\;\;\;\;\;\;\;\;\;\;\;\; \forall n \in \mathcal{P},\;\;\;\; \forall m \in \mathcal{N} \backslash n, \label{Eq.12_3}\\[2mm]
& p_{n,m} \leq 0, \;\;\;\;\;\;\;\;\;\;\;\;\;\;\;\; \forall n \in \mathcal{C},\;\;\;\; \forall m \in \mathcal{N} \backslash n. \label{Eq.12_4}
\end{align}
\end{subequations}
In second problem $\boldsymbol P_2$, the prosumers update the auxiliary variables $\boldsymbol \sigma_{n}$ based on the solution $p_{n,m}^{i+1}$ and constraint (\ref{Eq.9}) as:
\begin{subequations}\label{Eq.13}
\begin{align}
\min_{{\sigma_{n,m}}} \quad
&  \sum_{m\in \mathcal{N} \backslash n} \bigg \{ \pi_{n,m}^i ( \sigma_{n,m} - p_{n,m}^{i+1})  +  \frac{\kappa}{2} (\sigma_{n,m} - p_{n,m}^{i+1})^2 \bigg \}, \label{Eq.10_1}\\
\text{s.t.}\quad
& (\ref{Eq.9}).
\end{align}
\end{subequations}
It can be noticed that the auxiliary variables are only coupled between each pair of trading prosumers $n$ and $m$. Therefore, the optimization problem can be reformulated as:
\begin{subequations} \label{Eq.14}
    \begin{align}
    & \min_{{\sigma_{n,m},\sigma_{m,n}}}
      \sum_{m\in \mathcal{N} \backslash n} \bigg\{ \pi_{n,m} ( \sigma_{n,m} - p_{n,m}^{i+1})    +  \frac{\kappa}{2} \big(\sigma_{n,m} - p_{n,m}^{i+1} \big)^2 \bigg \} \nonumber \\
    & \quad \quad + \bigg \{\pi_{m,n} ( \sigma_{m,n} - p_{m,n}^{i+1})  +  \frac{\kappa}{2} ( \sigma_{m,n} - p_{m,n}^{i+1} )^2 \bigg \}.\\
    & \text{s.t.}\quad (\ref{Eq.9}).
    \end{align}
\end{subequations}
Based on $\boldsymbol p_n^{i+1}$ and $\boldsymbol \sigma_n^{i+1}$ obtained from solving the problems $\boldsymbol P_1$ and $\boldsymbol P_2$, the third problem $\boldsymbol P_3$ involves updating the the dual variable $\boldsymbol \pi_n$ as:
    \begin{equation} \label{Eq.15}
     \pi_{n,m}^{i+1} = \pi_{n,m}^{i} + \kappa ( \sigma_{n,m}^{i+1} - p_{n,m}^{i+1}).
    \end{equation}

\subsection{Closed-form Solution}
\subsubsection{Solution to the problem $\boldsymbol P_1$}
The optimization problem $\boldsymbol P_1$ in (\ref{Eq.12}) is a second-order quadratic problem that can be solved using Lagrangian based approach. In this case, the Lagrangian equation can be written as:
    \begin{equation} \label{Eq.16}
    \begin{split}
     \mathcal{L}(\hat{\boldsymbol p_n}, \boldsymbol \pi_n, \boldsymbol \tau_{n}, \boldsymbol \phi_{n}) =  c_n(\hat{p}_n) + \pi_{n} (\sigma_{n,m}^i - p_{n,m}) + \\ \frac{\kappa}{2} \sum_{m \in \mathcal{N} \backslash n}  \big(\sigma_{n,m}^i - p_{n,m} \big)^2 + {\tau_n^i} (\hat{p}_n - \hat{p}_n^{+}) - {\phi_n^i} (\hat{p}_n^{-} - \hat{p}_n),
     \end{split}
    \end{equation}
where $\tau_n$ and $\phi_n$ are the local Lagrangian parameters that define the boundary for the maximum and minimum power set point for prosumer $n$. With this definition, the first-order optimality conditions of the relaxed problem, given by the KKT conditions, are for all the prosumers as:
    \begin{equation} \label{Eq.17}
     2\alpha_n p_{n,m} + \beta_n - \pi_{n,m} - \kappa(\sigma_{n,m}^i - p_{n,m}) + \tau_n - \phi_n = 0,
    \end{equation}
which gives the closed-form solution as:
    \begin{equation} \label{Eq.18}
     p_{n,m}^{i+1} = \frac{\pi_{n,m}^{i}  - \tau_n^{i} + \phi_n^{i} -  \beta_n + \kappa \sigma_{n,m}^i}{2\alpha_n + \kappa}.
    \end{equation}
Considering the constraint (\ref{Eq.12_3}) and (\ref{Eq.12_4}), the complete solution can be derived as:
    \begin{equation} \label{Eq.19}
     p_{n,m}^{i+1} = \text{max}\big\{0,p_{n,m}^{i+1}\big\}, \;\;\;\;\;\;\;\; \forall n \in \mathcal{P},\;\;\;\; \forall m \in \mathcal{N} \backslash n,
    \end{equation}
    \begin{equation} \label{Eq.20}
     p_{n,m}^{i+1} = \text{min}\big\{0,p_{n,m}^{i+1}\big\}, \;\;\;\;\;\;\;\; \forall n \in \mathcal{C},\;\;\;\; \forall m \in \mathcal{N} \backslash n.
    \end{equation}
In the next step, local Lagrangian parameter $\tau$ and $\phi$ should be updated taking into the account complementary slackness as:
    \begin{equation} \label{Eq.21}
    \tau_n^{i+1} = \text{max}\big\{0,\tau_n^{i} + \rho \big(\hat{p}_n^{i+1} - \hat{p}_n^{+} \big) \big\},
    \end{equation}
    \begin{equation} \label{Eq.22}
    \phi_n^{i+1} = \text{max}\big\{0,\phi_n^{i} + \rho \big(\hat{p}_n^{-} - \hat{p}_n^{i+1} \big) \big\}.
    \end{equation}
where $\rho$ is a well-defined positive tuning parameter.

\subsubsection{Solution to the problem $\boldsymbol P_2$}
By applying Lagrangian approach with first-order KKT conditions in (\ref{Eq.14}), we can obtain the closed form solution for updating $\sigma_{n,m}$ as:
    \begin{equation} \label{Eq.23}
     \sigma_{n,m}^{i+1} = \frac{\kappa(p_{n,m}^{i+1} - p_{m,n}^{i+1}) - (\pi_{n,m}^{i} - \pi_{m,n}^{i})}{2\kappa}.
    \end{equation}
It should be noted that, the paired prosumers $n$ and $m$ share $\boldsymbol p$, $\boldsymbol \pi$, and $\boldsymbol \sigma$ in each iteration. The prosumers do not share total power generation or consumption profile but a partial quantity only; therefore, the privacy of the prosumers is not violated. The pseudo-code of the proposed P2P energy trading scheme is shown in Algorithm \ref{Algorithm_1}.
%
%\subsection{Convergence Error and Stopping Criterion}
The convergence error for the proposed algorithm can be calculated as:
\begin{equation}\label{Eq.24}
\delta^{i} =  \sqrt{{\delta_1^{i}}^2 + {\delta_2^{i}}^2},
\end{equation}
where $\delta_1^{i} = \big\| p_n^{i} - p_n^{i-1} \big\|_2$ and $\delta_2^{i} = \big\| \pi_n^{i} - \pi_n^{i-1} \big\|_2$. $\big\|A\|_2$ represents the $L_2$ norm of $A$. In this paper, the algorithm stopping condition was considered as $\delta \leq 10^{-2}$.

\begin{algorithm}[t]
%\small
\SetAlgoLined
\caption{Proposed distributed dynamic pricing between prosumer \textit{n} and \textit{m} at prosumer-end \textit{n} using F-ADMM}\label{Algorithm_1}
%\SetKwData{Left}{left}\SetKwData{This}{this}\SetKwData{Up}{up}
\SetKwFunction{Union}{Union}\SetKwFunction{FindCompress}{FindCompress}
\SetKwInOut{Input}{Input}\SetKwInOut{Output}{Output}
\Input{$\{\delta, \pi_{n,m}^0, \tilde{\pi}_{n,m}^1, \sigma_{n,m}^1, \mu_n^1, i=1\}$}
\Output{$\{p_{n,m}^*, \tilde{\pi}_{n,m}^*\}$}
\BlankLine
%\For{$t=1:T$}{
\While{$\delta \leq 10^{-2}$}{
P1: Update $p_{n,m}^{i+1}$ as:
    \begin{equation*}
    p_{n,m}^{i+1} = \text{max}\bigg\{ 0,\frac{\tilde{\pi}_{n,m}^{i}  - \tau_n^{i} + \phi_n^{i} -  \beta_n + \kappa \sigma_{n,m}^i}{2\alpha_n + \kappa}\bigg\}_{n \in \mathcal{P}}
    \end{equation*}
    \begin{equation*}
    p_{n,m}^{i+1} = \text{min}\bigg\{0,\frac{\tilde{\pi}_{n,m}^{i}  - \tau_n^{i} + \phi_n^{i} -  \beta_n + \kappa \sigma_{n,m}^i}{2\alpha_n + \kappa}\bigg\}_{n \in \mathcal{C}}
    \end{equation*}\\
P2: Update $\sigma_{n,m}^{i+1}$ as:
    \begin{equation*}
     \sigma_{n,m}^{i+1} = \frac{\kappa(p_{n,m}^{i+1} - p_{m,n}^{i+1}) - (\pi_{n,m}^{i} - \pi_{m,n}^{i})}{2\kappa}
    \end{equation*}\\
P3: Update $\tilde{\pi}_{n,m}^{i+1}$ as:
    \begin{equation*}
    \pi_{n,m}^{i} = \tilde{\pi}_{n,m}^i + \kappa (\sigma_{n,m}^{i+1} - p_{n,m}^{i+1})
    \end{equation*}
    \begin{equation*}
    \tilde{\pi}_{n,m}^{i+1} = \pi_{n,m}^i + \frac{\mu_n^i - 1}{\mu_n^{i+1}}(\pi_{n,m}^{i} - \pi_{n,m}^{i-1})
    \end{equation*}
where,  $\mu_n^{i+1} =\frac{1 + \sqrt{1+4{\mu_n^{i}}^2}}{2}$\\
%    \begin{equation*}
%    \mu_n^{i+1} = \frac{1 + \sqrt{1+4{\mu_n^{i+1}}^2}}{2}
%    \end{equation*}\\
Calculate convergence error $\delta$ using (\ref{Eq.24})\\
}
\scriptsize \% In each step of an iteration, transmit and receive variables with prosumer $m$
%}
\SetAlgoLined
\end{algorithm}

\vspace{-8pt}
\section{Result and Analysis}\label{Result}
\begin{figure}[t]
    \centering
    \includegraphics[scale=0.75]{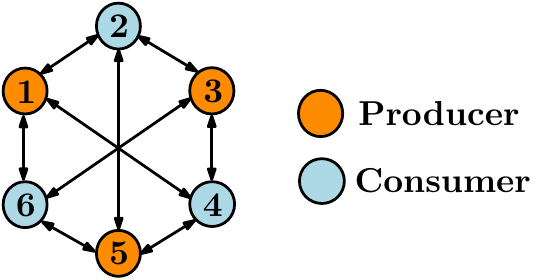}
    \caption{P2P communication benchmark in a distribution system.}
    \label{fig. 2}
    \vspace{-10pt}
    \end{figure}

This section presents the analysis of the simulation result obtained using the proposed scheme. A 6-prosumer based distributed system was considered, and the P2P connection is shown in Fig. \ref{fig. 2}, where the prosumer indexed with odd and even numbers are considered as an energy producer and consumer, respectively. Although the result was obtained only for one hour, the proposed dynamic pricing strategy can readily be applied for multi-time step systems.
\begin{table}[htbp]
\scriptsize
\caption{Technical parameters for sellers.}
\centering
\begin{tabular}{p{0.75cm} p{1.20cm} p{1cm} p{1cm} p{0.9cm} p{0.9cm} p{0.9cm}} %
\hline
\hline
Prosumer ID & $\alpha$ (\textcent/$kWh^2$) & $\beta$ (\textcent/kWh) & $\gamma$ (\textcent)  & $\hat{p}_n^+$ (kW) & $\hat{p}_n^-$ (kW)\\ [0.5ex]
\hline
1  & 0.455 &	2.275   & 0    &     28   & 10  \\
2  & 0.975 &	14.69   & 0    &	 -20  & -30  \\
3  & 0.520 &	2.600   & 0    &	 30   & 10  \\
4  & 0.884 &	14.95   & 0    &     -14  & -24  \\
5  & 0.585 &	3.770   & 0    &	 40   & 10  \\
6  & 0.676 &	15.275  & 0    &	 -10  & -34  \\
%[1ex] % [1ex] adds vertical space
\hline
\hline
\end{tabular}
\label{tab:1}
\vspace{-8pt}
\end{table}
Prosumer's technical data is presented in Table \ref{tab:1}. The energy selling and buying price to and from the grid was considered as 5\textcent/kWh and 15\textcent/kWh, respectively. Therefore, the settled price signal between the prosumers should be at this rate to fully utilize the advantages of the P2P market. The local optimization tuning parameter $\rho = 0.25$, and penalty parameter $\kappa = 0.5$. All the optimization variables are initialized at 0. The proposed distributed optimization was performed with MATLAB 2018.

    \begin{figure}[t!]
    \centering
    \includegraphics[scale=0.55]{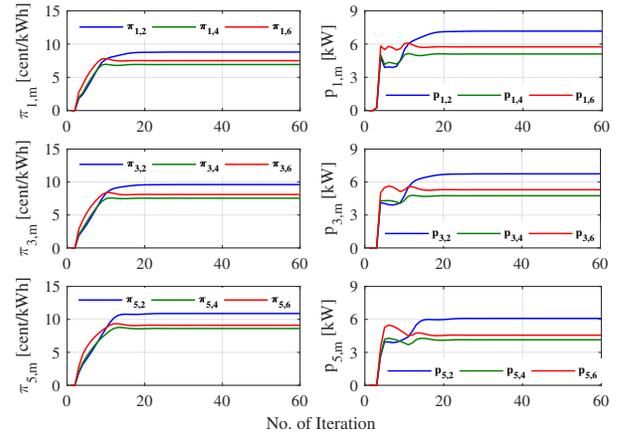}
    \caption{Evolution of price ($\pi_{n,m}$) and power ($p_{n,m}$) estimated by the producers.}
    \label{fig. 3}
    \vspace{-10pt}
    \end{figure}
    \begin{figure}[t!]
    \centering
    \includegraphics[scale=0.55]{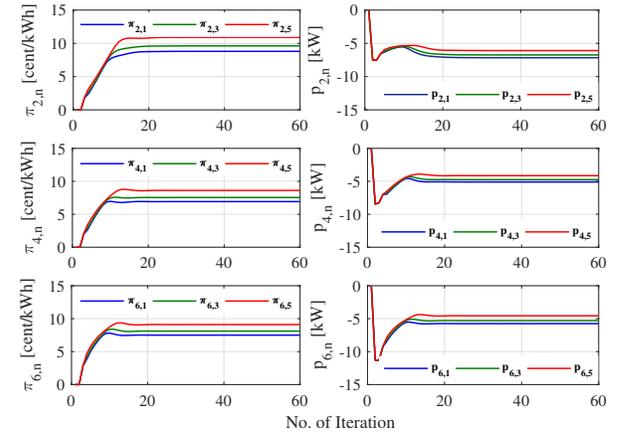}
    \caption{Evolution of price ($\pi_{m,n}$) and power ($p_{m,n}$) estimated by the consumers.}
    \label{fig. 4}
    \vspace{-6pt}
    \end{figure}

The evolution of the traded power and price between the prosumers are shown in Fig. \ref{fig. 3} and \ref{fig. 4}. In can be seen that the prosumers reach the steady-state of the trade after a finite number of iteration, which is less than 30. In the proposed scheme, the quantities of traded power and the price mainly depend on the prosumer's utility function coefficients $\alpha$ and $\beta$, which can be noticed from obtained results. It can be seen from Fig. \ref{fig. 3} that the estimated price of \textit{producer 1} is comparatively lower than the \textit{producer 3} and \textit{5}; consequently, \textit{producer 1} sells more power than the others, which is 18.1068 kW. The settled prices and power quantities are presented in Table \ref{tab:2} and \ref{tab:3}. Contrarily, the \textit{consumer 2} buys more power due to willingness to pay more price than the other consumers. The total traded power in the system is 48.0805 kW.

\begin{table}[t!]
\scriptsize
\caption{Settled price, $\pi_{n,m}$ (\textcent/kWh).}
\centering
\begin{tabular}{| p{0.5cm} | p{1.00cm} | p{1cm} | p{1cm} |} %
\hline
      $\mathcal{N}$ & \textbf{1} & \textbf{3} & \textbf{5}\\
      \hline
      \textbf{2} &  8.7956 & 9.6151 & 9.1074\\ \hline
      \textbf{4} &  6.9198 & 7.5422 & 8.6181\\ \hline
      \textbf{6} &  7.5049 & 8.1109 & 9.1074\\
\hline
\end{tabular}
\label{tab:2}
\vspace{-4pt}
\end{table}

\begin{table}[t!]
\scriptsize
\caption{Traded power, $|p_{n,m}|$ (kW).}
\centering
\begin{tabular}{| p{0.5cm} | p{1.00cm} | p{1cm} | p{1cm} | p{1cm}|} %
      \hline
      $\mathcal{N}$ & \textbf{1} & \textbf{3} & \textbf{5} & \textbf{Total}\\
      \hline
      \textbf{2} & 7.1655  & 6.7453  & 4.5619 & 18.4726\\ \hline
      \textbf{4} & 5.1042  & 4.7522  & 4.1436 & 14.0000 \\ \hline
      \textbf{6} & 5.7471  & 5.2989  & 4.5619 & 15.6079 \\  \hline
      \textbf{Total} &  18.0168 & 16.7963 & 13.2674 & \textbf{48.0805}\\
      \hline
\end{tabular}
\label{tab:3}
\vspace{-14pt}
\end{table}

\begin{figure}[t] % "[t!]" placement specifier just for this example
\begin{subfigure}{0.25\textwidth}
\includegraphics[width=\linewidth]{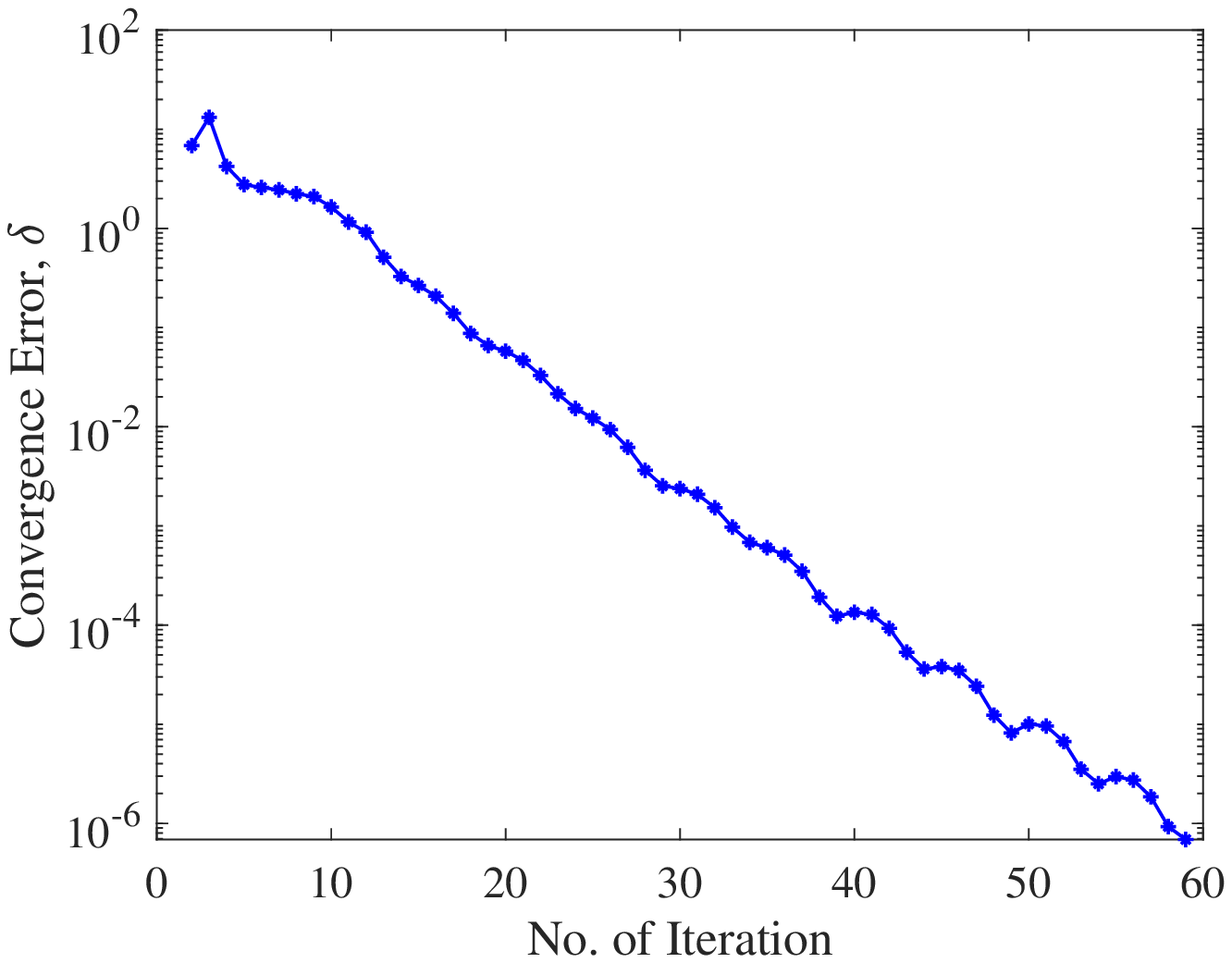}
\caption{} \label{fig 5:sfig1}
\end{subfigure}%\hspace*{\fill}
\begin{subfigure}{0.25\textwidth}
\includegraphics[width=\linewidth]{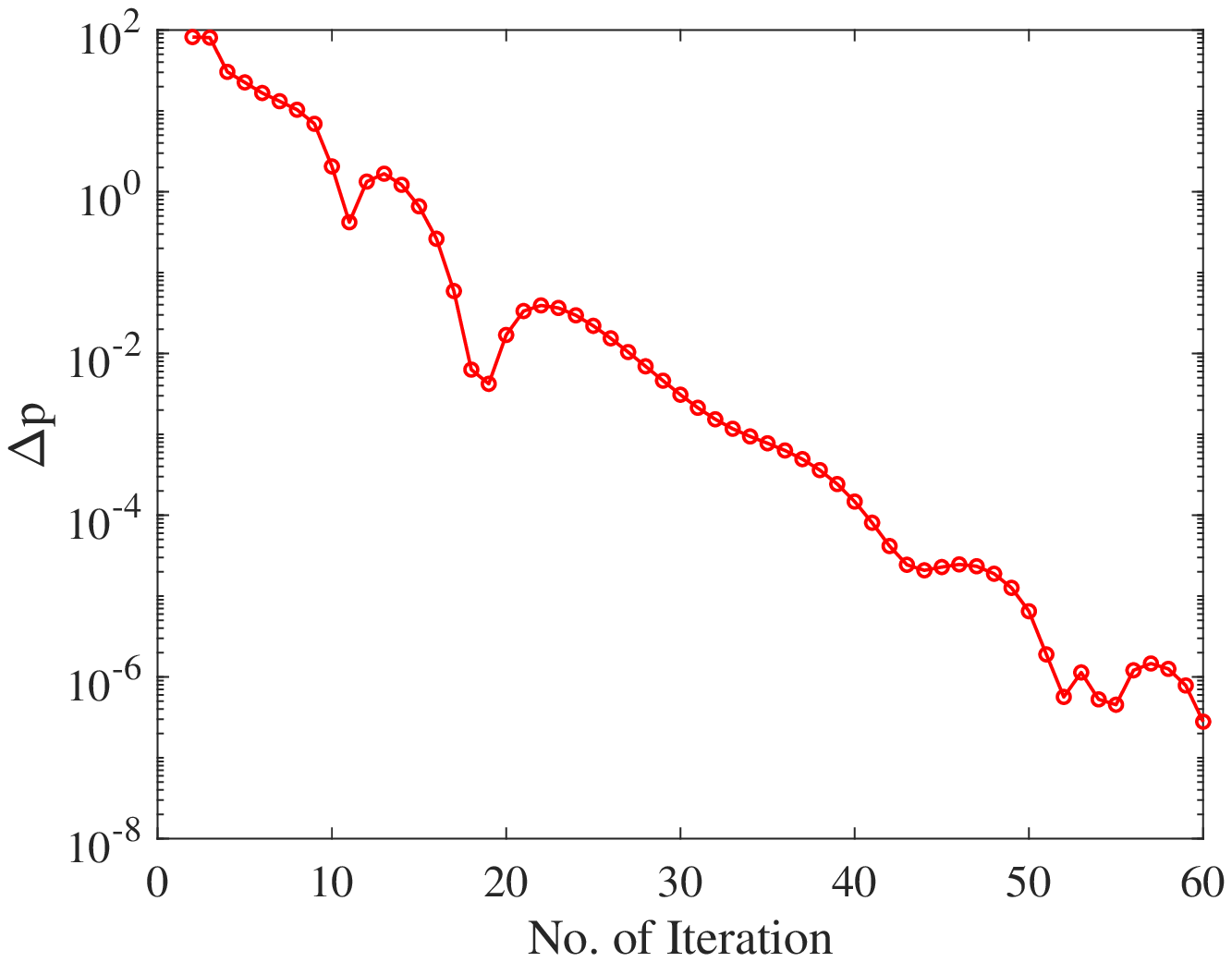}
\caption{} \label{fig 5:sfig2}
\end{subfigure}
\caption{Evolution of convergence error and power mismatch in the system.}
\label{fig. 5}
\vspace{-18pt}
\end{figure}

The convergence error of the proposed algorithm was calculated using (\ref{Eq.24}), which is shown in Fig \ref{fig 5:sfig1}. It can be seen that the algorithm converges very fast in less than 30 iterations with convergence condition of $\delta \leq 10^{-2}$. When the algorithm converges, the global power balance should be satisfied as well, which implies, $\Delta p = \sum_{n \in \mathcal{N}} \hat{p}_n = 0$. The evolution of the power balance is shown in Fig. \ref{fig 5:sfig2}. It can be noticed that when the algorithm reaches the convergence, the global power mismatch is close to zero. The computation time to reach the convergence is 0.122 s, and it takes 26 iterations.

%%
%    \begin{figure}[t!]
%    \centering
%    \includegraphics[scale=0.4]{deltap.eps}
%    \caption{Power mismatch in the overall system.}
%    \label{fig. 4}
%    \vspace{-6pt}
%    \end{figure}
%
%%
%    \begin{figure}[t!]
%    \centering
%    \includegraphics[scale=0.4]{conv_error.eps}
%    \caption{Convergence error in the proposed algorithm.}
%    \label{fig. 4}
%    \vspace{-6pt}
%    \end{figure}

\vspace{-8pt}
\section{Conclusion}\label{conclusion}
In this paper, a dynamic pricing strategy was proposed for prosumer centric P2P transactive energy systems in smart grid using a F-ADMM-based algorithm. The closed-form solution was derived to expedite the computation time. In the proposed scheme, no central coordinator is required and the prosumers can settle the market with sharing minimal information among themselves. The proposed schemes provide feasible solution and the convergence was verified through numerical results. However, the line congestion was not considered in this paper which is the future work. Moreover, a JAVA-based multi-agent system will be developed for P2P communication.
\vspace{-8pt}
\section*{Acknowledgement}
This work was supported by the National Science Foundation grant ECCS-1554626.
\vspace{-8pt}
\bibliographystyle{IEEEtran}
\bibliography{references}

\begin{thebibliography}{10}
\providecommand{\url}[1]{#1}
\csname url@samestyle\endcsname
\providecommand{\newblock}{\relax}
\providecommand{\bibinfo}[2]{#2}
\providecommand{\BIBentrySTDinterwordspacing}{\spaceskip=0pt\relax}
\providecommand{\BIBentryALTinterwordstretchfactor}{4}
\providecommand{\BIBentryALTinterwordspacing}{\spaceskip=\fontdimen2\font plus
\BIBentryALTinterwordstretchfactor\fontdimen3\font minus
  \fontdimen4\font\relax}
\providecommand{\BIBforeignlanguage}[2]{{%
\expandafter\ifx\csname l@#1\endcsname\relax
\typeout{** WARNING: IEEEtran.bst: No hyphenation pattern has been}%
\typeout{** loaded for the language `#1'. Using the pattern for}%
\typeout{** the default language instead.}%
\else
\language=\csname l@#1\endcsname
\fi
#2}}
\providecommand{\BIBdecl}{\relax}
\BIBdecl

\bibitem{peck2017energy}
M.~E. Peck and D.~Wagman, ``Energy trading for fun and profit buy your
  neighbor's rooftop solar power or sell your own-it'll all be on a
  blockchain,'' \emph{IEEE Spectrum}, vol.~54, no.~10, pp. 56--61, 2017.

\bibitem{darghouth2011impact}
N.~R. Darghouth, G.~Barbose, and R.~Wiser, ``The impact of rate design and net
  metering on the bill savings from distributed pv for residential customers in
  california,'' \emph{Energy Policy}, vol.~39, no.~9, pp. 5243--5253, 2011.

\bibitem{couture2010policymaker}
T.~D. Couture, K.~Cory, C.~Kreycik, and E.~Williams, ``Policymaker's guide to
  feed-in tariff policy design,'' National Renewable Energy Lab.(NREL), Golden,
  CO (United States), Tech. Rep., 2010.

\bibitem{tushar2018peer}
W.~Tushar, T.~K. Saha, C.~Yuen, P.~Liddell, R.~Bean, and H.~V. Poor,
  ``Peer-to-peer energy trading with sustainable user participation: A game
  theoretic approach,'' \emph{IEEE Access}, vol.~6, pp. 62\,932--62\,943, 2018.

\bibitem{morstyn2018using}
T.~Morstyn, N.~Farrell, S.~J. Darby, and M.~D. McCulloch, ``Using peer-to-peer
  energy-trading platforms to incentivize prosumers to form federated power
  plants,'' \emph{Nature Energy}, vol.~3, no.~2, p.~94, 2018.

\bibitem{guerrero2018decentralized}
J.~Guerrero, A.~C. Chapman, and G.~Verbi{\v{c}}, ``Decentralized p2p energy
  trading under network constraints in a low-voltage network,'' \emph{IEEE
  Transactions on Smart Grid}, 2018.

\bibitem{wu2018user}
S.~Wu, F.~Zhang, and D.~Li, ``User-centric peer-to-peer energy trading
  mechanisms for residential microgrids,'' in \emph{2nd IEEE Conference on
  Energy Internet and Energy System Integration (EI2)}.\hskip 1em plus 0.5em
  minus 0.4em\relax IEEE, 2018, pp. 1--6.

\bibitem{long2017peer}
C.~Long, J.~Wu, C.~Zhang, L.~Thomas, M.~Cheng, and N.~Jenkins, ``Peer-to-peer
  energy trading in a community microgrid,'' in \emph{PES General
  Meeting,}.\hskip 1em plus 0.5em minus 0.4em\relax IEEE, 2017, pp. 1--5.

\bibitem{narayanan2018economic}
A.~Narayanan, J.~Haapaniemi, T.~Kaipia, and J.~Partanen, ``Economic impacts of
  power-based tariffs on peer-to-peer electricity exchange in community
  microgrids,'' in \emph{15th International Conference on the European Energy
  Market (EEM)}.\hskip 1em plus 0.5em minus 0.4em\relax IEEE, 2018, pp. 1--5.

\bibitem{khorasany2017auction}
M.~Khorasany, Y.~Mishra, and G.~Ledwich, ``Auction based energy trading in
  transactive energy market with active participation of prosumers and
  consumers,'' in \emph{Australasian Universities Power Engineering Conference
  (AUPEC)}.\hskip 1em plus 0.5em minus 0.4em\relax IEEE, 2017, pp. 1--6.

\bibitem{peerkhorasany2017peer}
------, ``Peer-to-peer market clearing framework for ders using knapsack
  approximation algorithm,'' in \emph{IEEE PES Innovative Smart Grid
  Technologies Conference Europe (ISGT-Europe)}, 2017, pp. 1--6.

\bibitem{zhang2019two}
Z.~Zhang, H.~Tang, Q.~Huang, and W.-J. Lee, ``Two-stages bidding strategies for
  residential microgrids based peer-to-peer energy trading,'' in \emph{IEEE/IAS
  55th Industrial and Commercial Power Systems Technical Conference
  (I\&CPS)}.\hskip 1em plus 0.5em minus 0.4em\relax IEEE, 2019, pp. 1--9.

\bibitem{liu2015energy}
T.~Liu, X.~Tan, B.~Sun, Y.~Wu, X.~Guan, and D.~H. Tsang, ``Energy management of
  cooperative microgrids with p2p energy sharing in distribution networks,'' in
  \emph{IEEE international conference on smart grid communications
  (SmartGridComm)}.\hskip 1em plus 0.5em minus 0.4em\relax IEEE, 2015, pp.
  410--415.

\bibitem{yoo2017peer}
Y.-S. Yoo, T.~Hwang, S.~Kang, S.~S. Newaz, I.-W. Lee, and J.~K. Choi,
  ``Peer-to-peer based energy trading system for heterogeneous small-scale
  ders,'' in \emph{International Conference on Information and Communication
  Technology Convergence (ICTC)}.\hskip 1em plus 0.5em minus 0.4em\relax IEEE,
  2017, pp. 813--816.

\bibitem{moret2018negotiation}
F.~Moret, T.~Baroche, E.~Sorin, and P.~Pinson, ``Negotiation algorithms for
  peer-to-peer electricity markets: Computational properties,'' in \emph{2018
  Power Systems Computation Conference (PSCC)}.\hskip 1em plus 0.5em minus
  0.4em\relax IEEE, 2018, pp. 1--7.

\bibitem{sorin2018consensus}
E.~Sorin, L.~Bobo, and P.~Pinson, ``Consensus-based approach to peer-to-peer
  electricity markets with product differentiation,'' \emph{IEEE Transactions
  on Power Systems}, vol.~34, no.~2, pp. 994--1004, 2018.

\bibitem{morstyn2018multi}
T.~Morstyn and M.~McCulloch, ``Multi-class energy management for peer-to-peer
  energy trading driven by prosumer preferences,'' \emph{IEEE Transactions on
  Power Systems}, 2018.

\bibitem{le2019peer}
H.~Le~Cadre, P.~Jacquot, C.~Wan, and C.~Alasseur, ``Peer-to-peer electricity
  market analysis: From variational to generalized nash equilibrium,''
  \emph{European Journal of Operational Research}, 2019.

\bibitem{anoh2019energy}
K.~Anoh, S.~Maharjan, A.~Ikpehai, Y.~Zhang, and B.~Adebisi, ``Energy
  peer-to-peer trading in virtual microgrids in smart grids: A game-theoretic
  approach,'' \emph{IEEE Transactions on Smart Grid}, 2019.

\bibitem{goldstein2014fast}
T.~Goldstein, B.~O'Donoghue, S.~Setzer, and R.~Baraniuk, ``Fast alternating
  direction optimization methods,'' \emph{SIAM Journal on Imaging Sciences},
  vol.~7, no.~3, pp. 1588--1623, 2014.

\end{thebibliography}

\end{document}